\documentclass[preprint,1p,12pt]{elsarticle}

\biboptions{authoryear}

\usepackage{epsfig}
\usepackage{float}

\usepackage{amssymb}
\usepackage{amsmath}

\newcommand{\apj}{    {\it The Astrophys. J.}}

\newcommand{\solphys}{{\it Solar Phys.}}

\usepackage[colorlinks = true, linkcolor = blue, urlcolor  = blue, citecolor = blue, anchorcolor = blue]{hyperref}
\usepackage{lineno}

\journal{Advances in Space Research}

\begin{document}

\begin{frontmatter}
\title{An application of the weighted horizontal magnetic gradient to solar compact and eruptive events}
\author{M. B. Kors\'os$^{1,2}$}
\author{Michael S. Ruderman$^{1}$}
\author{R. Erd\'elyi$^{1,3}$}
\address{1.Solar Physics \& Space Plasma Research Center (SP2RC), University of Sheffield, Hounsfield Road, S3 7RH, UK}
\address{2.Debrecen Heliophysical Observatory (DHO), Konkoly Astronomical Institute, Research Centre for Astronomy and Earth Sciences, Hungarian Academy of Sciences, Debrecen, P.O.Box 30, H-4010, Hungary}
\address{3.Department of Astronomy, E\"otv\"os Lor\'and University, Budapest, Hungary}

\begin{abstract}
We propose to apply the weighted horizontal magnetic gradient ($WG_{M}$), introduced in Korsos et al (2015), for analysing the pre-flare and pre-CME behaviour and evolution of Active Regions (ARs) using the SDO/HMI-Debrecen Data catalogue. To demonstrate the power of investigative capabilities of the $WG_M$ method, in terms of flare and CME eruptions, we studied two typical ARs, namely, AR 12158 and AR 12192. The choice of ARs represent canonical cases. AR 12158 produced an X1.6 flare with fast ``halo" CME ($v_{linear}$=1267 $km s^{-1}$) while in AR 12192 there  occurred a range of powerful X-class eruptions, i.e. X1.1, X1.6, X3.1, X1.0, X2.0 and X2.0-class energetic flares, interestingly, none with an accompanying CME. The value itself and temporal variation of $WG_{M}$ is found to possess  potentially important diagnostic information about the intensity of the expected flare class. Furthermore, we have also estimated the flare onset time from the relationship of duration of converging and diverging motions of the area-weighted barycenters of two subgroups of opposite magnetic polarities.
This test turns out not only to provide information about the intensity of the expected flare-class and the flare onset time but may also indicate whether a flare will occur with/without fast CME.
 We have also found that, in the case when the negative polarity barycenter has moved around and the positive one ``remained" at the same coordinates preceding eruption, the flare occurred with fast ``halo" CME.
 Otherwise, when both the negative and the positive polarity barycenters have moved around, the AR produced flares without CME. If these properties found for the movement of the barycenters are generic pre-cursors of CME eruption (or lack of it), identifying them may serve as an excellent pre-condition for refining the forecast of the lift-off of CMEs. 
\end{abstract}


\end{frontmatter}

\section{Introduction}

  There are many kinds of eruptions on the Sun and from these the solar flares and coronal mass ejections (CMEs) are the most gigantic energy explosions. These two major eruptions are powered by the free energy stored in the stressed magnetic fields in active regions (ARs). Sunspots appear as dark spots compared to surrounding regions on the photosphere  and are considered as good markers of ARs. The concentration of magnetic field fluxes of AR, often modelled as flux tubes, reduce the temperature in the photosphere by inhibiting convection. Strongly twisted magnetic flux tubes and strongly sheared magnetic structures are candidates for facilitating the high intensity flares and flux rope eruption from AR. A number of specific mechanisms are proposed to lead to flare and CME occurrences, e.g. sunspot rotation \citep{Yan2007, Zhang2007, Yan2009, Chandra2011, Hardersen2011,Vemareddy2016} and shearing motion of the sunspots at photosphere \citep{Vemareddy2012} which contribute to helicity and accumulation of magnetic energy of an AR \citep{Torok2003, Demoulin2007, Demoulin2009}. The magnetically complicated and highly dynamic delta-type sunspot groups are more likely for flare and CME genesis than bipolar ARs, see e.g. \cite{Kunzel1960}, \cite{Sammis2000}. It is now also well known that solar flares and CMEs occur close to the polarity inversion line (PIL) \citep{Louis2015}. The PIL can be defined as the boundary separating positive and negative magnetic polarities \citep{Babcock1955}.
  
       Flare and CME often accompany each other, but not always. \cite{Yashiro2006} found that the probability of a low energetic flare with CME occurrence  is much smaller than an intensive flare being associated with a large CME. If these two phenomena do occur together then the pre-, rise- or decay-phase of a flare is temporally associate with the initial-, impulsive acceleration- or propagation-phase of a CME \citep{Zhang2001}. 
  
  A main difference between solar flare and CME is the scale on which they occur. A flare is small and more local compared to a CME. Flares occur mainly in the low solar atmosphere where magnetic field lines of an AR are concentrated. CME is, however, an absolutely massive eruption that may occur on very large scales. A CME, in terms of its developed size, can even be bigger than the Sun itself. Furthermore, a flare evolves more rapidly and produces radiation at various wavelengths which may have a nasty consequence. For example, the higher frequencies can reach the Earth in about 8 mins causing telecommunication disruption. Flares and CMEs are, however, likely parts of a single, magnetically-driven physical phenomenon, called magnetic reconnection, where highly fluctuating magnetic fields collapse to form a lower-energy state. The magnetic shockwave generated by a CME usually may reach the Earth in 18 to 36 hours the Earth if it has the earthward propagation direction. 
  
  Most of the electromagnetic energy of a flare is spread over frequencies \citep{Lin2003} outside of the visible, e.g. in x- and $\gamma$-range. Therefore the majority of flares must be observed with instruments capable of measurements in x- and $\gamma$- wavelength ranges, as e.g. the Geostationary Operational Environmental Satellite (GOES). Measurements of the maximum flux of x-ray at wavelengths from 0.1 to 0.8 nm near Earth, as taken by the XRS instrument on-board the GOES-15 satellite, are classed as A, B, C, M, or X type flares, where the series of GOES satellites operate back from 1975. These five classes of flare intensity categories are further divided, on a logarithmic scale, labelled from 1 to 9. The medium category of solar flare classification is the M-class flare that may cause smaller or occasionally more serious radio blackouts. The X-intensity flares may give rise from strong to extreme radio blackouts on the daylight side of the Earth.  
  
  The measured velocity of CME generally is the radial propagation speed of the upper part of a CME frontal loop. If the linear speed is between 500-800 $km s^{-1}$ then it is called slow CME, but when the linear velocity is over 800 $km s^{-1}$ than it is referred to as fast CME \citep{Ying2016}.  \cite{Alicia 2011}  found that ``halo" CMEs are associated with the most energetic flares. The Earth-directed ``halo" CMEs are capable of causing very strong geomagnetic storms, therefore, their prediction has more interest in general \citep{Chen2011}. \cite{Zhang2003} found that the flare association with fast CME tends to happen within half an hour of the CME onset. Otherwise, \cite{Zhang2003} presented that the relationship of flare associated with slow CME onsets is less correlated.

Flare and CME have a major impact on our life and our technological systems. An intense flare may ionise the upper atmosphere of the Earth which then will block the radio signals and disrupt radio communication. The shock wave of a propagating CME may cause a geomagnetic storm in the Earth's magnetosphere. CMEs can have a major effect on modern society's way of life. Very energetic particles can cause radiation poisoning to organic systems in space. The charged particles can also disrupt satellites and even our telecommunication and GPS-based navigation systems may be affected seriously. Magnetic storms may occasionally have major impact on our power grids and pipe lines on Earth. A few very strong magnetic storms have been known to black out entire regions (e.g. Canada Quebec was catastrophically affected on March 10, 1989). Therefore improving flare forecast and CME predictability along with understanding the underlying physics is of paramount importance. 
 
 Here, we investigate the evolution of opposite magnetic polarities near the PIL of two ARs, namely, AR 12158 and AR 12192. The AR 12158 produced an X1.6 energetic flare with a fast  ``halo" CME and the AR 12192 was a very intense flare-producer. During perceptibility of AR 12192 it was a cradle to five X-class flares without a single (known) CME. In Section \ref{tool}, we present briefly the concept of the weighted horizontal magnetic gradient ($WG_{M}$) method proposed by \cite{Korsos2015}. In Section \ref{Analyses}, we outline our detailed analysis of the two ARs  and summarise our findings  based on applying the $WG_{M}$ tool. Finally, we provide discussions of our results and draw conclusions in Sections \ref{conclusion}.

 \section{Tools of the Analysis} \label{tool}
 
 \subsection{Analysis with weighted horizontal magnetic gradient method}
 
\cite{Korsos2015} introduced the weighted horizontal magnetic gradient (denoted as $WG_{M}$) between two opposite magnetic polarity sunspot groups, and demonstrated that $WG_{M}$ could be applied to forecast the flare energy and the onset time of solar flare-class above M5. The distinguishing pre-flare behaviour of $WG_{M}$ is that it has a steep rise and a high maximum value followed by a less steep decrease which ends with flare(s). Note that the flare does not occur at the moment of reaching the maximum value of $WG_{M}$, but afterwards during it descending phase. 

The first important diagnostic information is the intensity of expected flares (let us denote it by $S$) obtainable from the maximum value of the $WG_{M}$ according to:

\begin{equation}
S_{flare} = a \cdot WG^{max}_{M} + b,  
\label{intensity}
\end{equation}
where $a =3.58\cdot 10^{-11} \pm 0.4\cdot 10^{-11}$ W/(m$\cdot$Wb)
and  $b =0.08 \cdot 10^{-5} \pm 1.38\cdot 10^{-5}$ W/m$^{2}$. The standard error is $ \pm 3 \cdot 10^{-5}$ W/m$^{2}$ \citep{Korsos2015, Korsos2016}.
 
Furthermore, the definition of $WG_{M}$ contains two components: total unsigned magnetic flux  and the distance between the area-weighted barycenters of spot groups of opposite polarities. The second potentially important diagnostic information is the connection between the duration of converging-diverging motion of the area-weighted barycenters of opposite polarities that seem to be indicative of the next flare(s) for {\it all} cases we investigated in \cite{Korsos2015}. The prediction of the flare onset time is based on the relationship found  between the duration of diverging motion of the area-weighted centres of opposite polarities until the flare onset ($T_{D+F}$) and time of the compressing motion ($T_{C}$) of area-weighted centres of the opposite polarities. \cite{Korsos2015} have classified the selected spot groups of their study by age - into younger  or older than three days - and repeated the investigation separately for these two groups, in order to determine how fundamental this relationship may be. The following regression found may be one of the most useful results of the prediction method of the $WG_{M}$:

\begin{equation}
T_{pred} = a_{1} \cdot T_{C} + b_{1},  
 \label{time}
\end{equation}

where $a_{1} =1.29 (0.85)$ [hr] and  $b_{1} =1.11 (12.8)$ [hr] in the younger (older) than three days case, respectively. Here, we note that the flare occurrences of the two investigated ARs manifested beyond the 72-hour threshold measured from the AR emergence at the photosphere. Therefore we use $a_{1} =0.85$ [hr] and $b_{1} =12.8$ [hr] in Equation \ref{time} for the flare onset time estimation. 

In brief, $WG_{M}$ may be interpreted as a proxy of the available non-potential (i.e., free) energy to be released in a spot group. This is because this parameter is an essential pre-cursor (but not sufficient) of flares. It is found a strong increase and a high peak of $WG_{M}$ is needed for flaring, during which the system relaxes to lower state of energy. In fact, following \cite{Korsos2015} we may conclude: if the maximum of the released energy may be over $\sim$54\% of the maximum of the accumulated (free) energy, no further energetic flare(s) can be expected; but, if the maximum of the released flare energy is less than about 42\%, further flares are more probable. These important properties of the $WG_{M}$ method were found for flare(s) in the studied cases, therefore, these features may serve as useful and practical flare watch alert tools.

 \section{Analysis} \label{Analyses}

We have selected two ARs that may demonstrate well typical behaviours of pre-flare phase, flaring with fast CME and flaring without CME at all. For the analysis we employed the SDO/HMI-Debrecen Data sunspot catalogue \citep{Baranyi2016}. HMIDD provides accurate and detailed position, area, and mean estimated magnetic field information for all observable sunspots and sunspot groups at an hourly basis from 2010 to the end of 2014. 

\begin{figure}[ht!]
  \centering
    \includegraphics[scale=1.2]{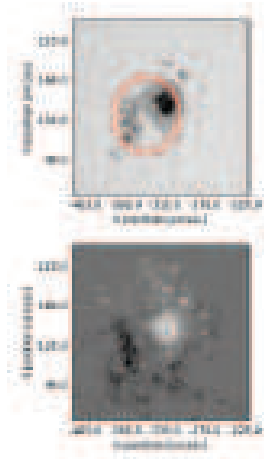}
    \caption{AR 12158. The X1.6 energetic class flare occurred with fast ``halo" CME ($v_{linear}=1267 km s^{-1}$).  Top panel: Intensity at 13:07 on 9 September 2014. Bottom panel: Magnetogram at 13:07 on 9 September 2014.}
    \label{12158pic}
\end{figure}

\begin{figure}[ht!]
 \centering
    \includegraphics[scale=1.2]{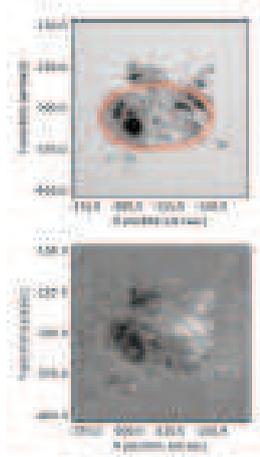} 
    \caption{AR 12192. The AR produced more X energetic class flares without any major CME. Top panel: Intensity at 13:07 on 22 October 2014. Bottom panel: Magnetogram at 13:07 on 22 October 2014.}
     \label{12192pic}
 
\end{figure}

Figure \ref{12158pic} shows AR 12158 and Figure \ref{12192pic} depicts AR 12192 in their white-light appearance (upper panel) and the corresponding magnetogram (bottom panel). The areas encircled by the red ellipses in the upper panel of Figures \ref{12158pic} and \ref{12192pic} are the study areas containing spots of opposite polarities. These study areas are where the most intense flares are in connection with the location of the strongest magnetic gradient.

   \begin{figure}[ht!]
     \centering
    \includegraphics{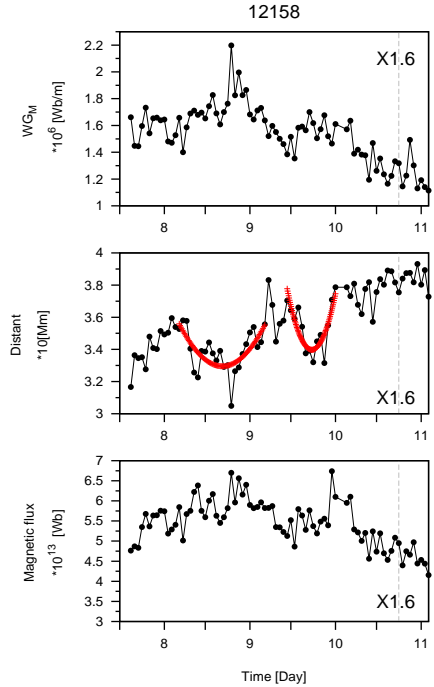}
    \caption{AR 12158. The X1.6 class energetic flare occurred with fast ``halo" CME ($v_{linear}=1267 km s^{-1}$).  Top panel: variation of $WG_{M}$ as a function of time; Middle panel: evolution of distance between the area-weighted barycenters of the spots of opposite polarities; Bottom panel: unsigned flux of all spots in the encircled area as a function of time.}
    \label{12158}
\end{figure}

   \begin{figure}[ht!]
     \centering
    \includegraphics{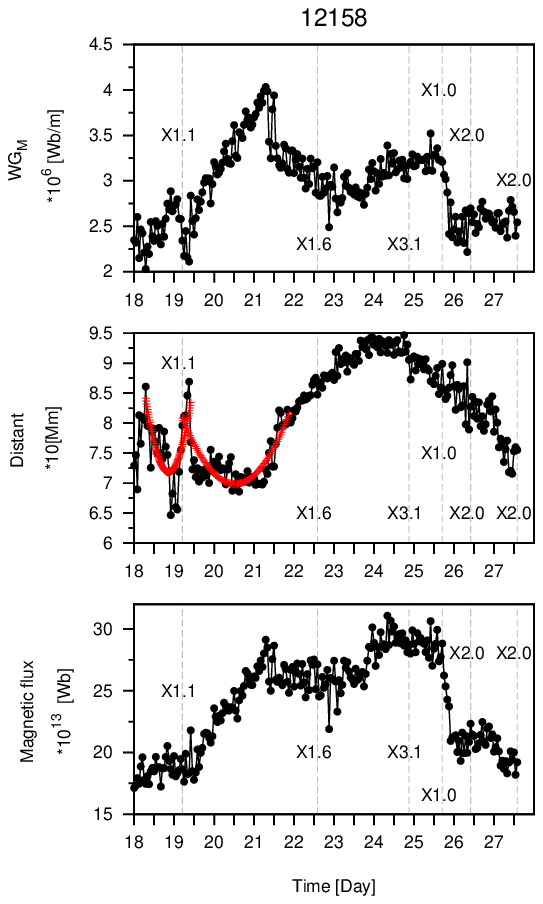} 
    \caption{AR 12192. The AR produced more X class energetic flares without major CME. Top panel: variation of $WG_{M}$ as a function of time; Middle panel: evolution of distance between the area-weighted barycenters of the spots of opposite polarities; Bottom panel: unsigned flux of all spots in the encircled area as a function of time.}
     \label{12192}
\end{figure}

The resulting diagrams of $WG_{M}$ analysis of AR 12158 are shown in Fig. \ref{12158}. AR 12158 is a typical example for the case of flare occurring with a fast CME \citep{Vemareddy2016,Zhang2017}. Fig. \ref{12158} shows the variation of the $WG_{M}$ value (top panel), distance (in the middle panel) and net magnetic flux (bottom panel). Preceding the X1.6-class flare, the $WG_{M}$ has increased to a maximum ($WG_{M}^{max}=2.19 \cdot 10^{6}$ Wb/m), after that, it is followed by a less steep decrease which ends with an X1.6-class energetic flare associated with a fast, ``halo" CME ($v_{linear}$=1267 $km s^{-1}$). At the same time of the increasing-decreasing phase of the $WG_M$, one can observe the converging and diverging motion of the area-weighted barycenters of two subgroups of opposite polarities in the selected cluster (highlighted with the red parabolae in the middle panel of Figure \ref{12158}). Furthermore, we can identify another converging and diverging motion before the  solar eruptions (X1.6 flare with fast CME).

When the maximum value of $WG_{M} =2.19 \cdot 10^{6}$ Wb/m is substituted into Equation (\ref{intensity}) that yields the predicted maximum flare intensity $S_{flare} = 8 \cdot 10^{-5}$ W/m$^{2}$  in the 1-8 \AA \, wavelength range of GOES, i.e. corresponding to an expected M8.0 flare, in apparent contrast to the X1.6 ($S_{flare} = 16 \cdot 10^{-5}$ W/m$^{2}$)  flare that actually took place. The highest intensity permitted by standard error would be X1.1, which is  in fact the same flare intensity-class that the actually measured X1.6 class intensity flare. Therefore, there seems to be an under-estimate. 

Let us now forecast the flare onset time. The accumulated free energy, represented by $WG_{M}$ as a proxy measure, is released in the form of flare and a fast ``halo" CME in the  case of AR 12158. Equation \ref{time} enables prediction of the onset time of the flare from the computed duration of the converging phase ($T_{C}$) of motion of opposite polarities. The first predicted onset time of the X1.6-class flare estimated from the first converging-diverging motion, highlighted with the first red parabola (middle panel, Fig. \ref{12158}), is 23.85 hrs after the minimum distance reached at 05:00 on 8 September 2014 according to Equation \ref{time}. This predicted onset time is actually somewhat far from the observed occurrence time of flare at 17:45 on 10 September 2014. However, the second predicted onset time of the X1.6-class flare from the second converging and diverging motion (second red parabola in middle panel of Figure \ref{12158}) is 18.75 hrs after the minimum distance reached at 18:00 on 9 September 2014 by Equation \ref{time}. This predicted onset time would be at 13:00 on 10 September 2014 which actually is now pretty close to 17:45 on 10 September 2014. Here, we conjecture that we may witness a failed flare eruption (first red parabola) followed by a belated true one. More similar cases would be needed to investigate before a firm conclusion is drawn. This, however, is beyond the scope of the current study.

Next, let us now investigate how $WG_{M}$ can be employed as a proxy for estimating the available non-potential energy to be released in a selected cluster. As described above in Section \ref{Analyses}: (i) if the maximum of the released energy is larger than 54\% no further energetic flare be expected, but (ii) if the maximum of the released flare energy is less than 42\%, further flaring is probable. Let us now calculate the required percentage from the relationship between the maximum value of $WG_{M}$ and value of $WG_{M}$ at the flare onset at photospheric level. The percentage computed for the X1.6 flare corresponds to a 40\% decrease meaning that we may expect further eruption(s) during the decreasing phase of $WG_{M}$. In fact, the X1.6 flare occurred with a huge coronal mass ejection which is a giant cloud of solar plasma drenched with magnetic field lines that are blown away from the Sun. After this major CME occurrence the magnetic topology of the AR 12158 is rearranged and seems to be stabilised.

Let us now turn to the case of AR 12192 (see Fig. \ref{12192}). This AR is a good example for the flare to occur without CME. It is fair to mention that this AR is rather unusual as being extremely large but, interestingly, CME-poor. AR 12192 has been described well in the literature \citep{Veronig2015, Jiang2016, Liu2016}. Here, again, we notice the following remarkable properties of the $WG_M$ and the distance.
First of all, in Fig. \ref{12192}, the steep rise and a high maximum value of the  weighted horizontal gradient of the magnetic field is still followed by a less steep decrease which ends with X1.1-class flare. Next, about 13 hrs later, after the first maximum value of the $WG_{M}$, one finds another steep rise and the associated high maximum value of the flux gradient, followed again, by a less steeper decrease which ends with the series of  X1.6, X3.1, X1.0, X2.0 and X2.0-class energetic flares.
The first maximum value of $WG_{M} = 2.9 \cdot 10^{6}$ Wb/m  gives, according to Eq. \ref{intensity}, the maximum predicted intensity $S_{flare} = 10 \cdot 10^{-5}$ W/m$^{2}$, i.e. corresponding to an expected X1.0 flare. This highest predicted intensity flare class is  in fact very close to the actually measured X1.1-class intensity flare. In this case, the accumulated free energy is released by the X1.1 flare. Next, we investigate the percentage of the decrease of the  $WG_{M}$ to the flare. This percentage is a mere 28\% when the X1.1-class flare occurred, so more flare(s) would be expected. The second maximum value of $WG_{M} = 4.0 \cdot 10^{6}$ Wb/m  gives, according to Eq. \ref{intensity}, the maximum predicted intensity $S_{flare} = 14 \cdot 10^{-5}$ W/m$^{2}$, i.e. corresponding to an expected X1.4 flare, in good agreement with the next X-class flare(s) that actually took place.

  \begin{figure}[ht!]
    \includegraphics{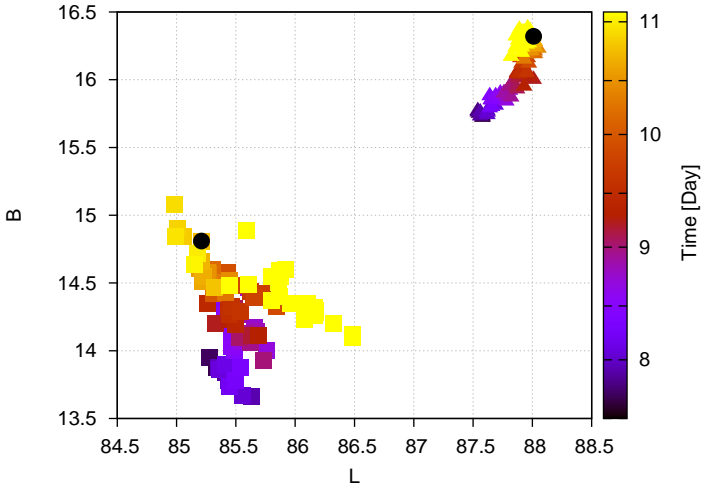}
    \caption{AR 12158. The area-weighted barycentre of the negative (square) and positive (triangle) polarity motion before the eruption in Carrington-coordinate system ({\it x} axis is longitude {\it L}; {\it y} axis is latitude {\it B}).The colour bar demonstrates the elapsed date [Day]. Black filled circles highlight the positions of the two barycentres of  polarities at the flare onset time.}
    \label{12158map}
  \end{figure}

   \begin{figure}[ht!]
    \includegraphics{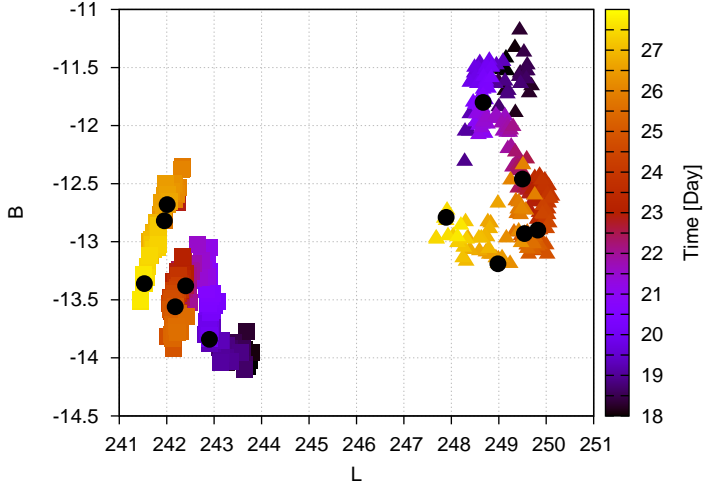} 
    \caption{Same as Fig. \ref{12158map} but for AR 12192.}
     \label{12192map}
\end{figure}

 Furthermore, we note that as we follow the evolution of the distance parameter in time we can clearly see the duration of converging and diverging motion of area-weighted barycenters of opposite polarities  (highlighted again with the red parabola in the middle panel of Figure \ref{12192}) before the X1.1- and the further duration of converging and diverging motion (the second red parabola in the middle panel of Figure \ref{12192}) for the first flare of the series of the subsequent X1.6, X3.1, X1.0, X2.0 and X2.0 class flares. In fact, according to the flare forecast theory based on observed sunspot data \citep[see][]{Korsos2015}, the predicted onset time of the X1.1 flare is 10 hrs earlier than the observered flare onset time. Next, we estimate the consecutive expected flare onset time from the approaching and receding phases of the distance before the flare series X1.6, X3.1, X1.0, X2.0 and X2.0-class energetic flares. The predicted onset time of the next flare is at 02:00 on 21 September 2014, according to Equation \ref{time}, and the first flare (X1.6) occurred 12 hrs later. Again, this may be a somewhat early prediction.  

In Figure \ref{12192}, we can see that after the second maximum of $WG_{M}$ during the decreasing phase more X-class flares happened. Now, we calculate the percentage of the decrease of the  $WG_{M}$ during the decreasing phase.
So, after the second maximum value of the $WG_{M}$, in the case of the X1.6-class flare, the percentage decrease is only 20\% and, indeed, more flares occurred as expected from the temporal evolution of $WG_{M}$. An X3.1-class flare appears after the X1.6-flare and the associated percentage of decreasing is 23\%. The further percentage is 20\% at the X1.0-class flare onset and the percentage is 35\% at the X2.0-class flare onset. The last observed flare is X2.0-class when the percentage of decreasing is 38\% at the flare onset but no more flares seem to be observed during this decreasing phase of the $WG_{M}$. Here, we note that AR 12192 turned out off the line of sight at the west limb, so we do not know whether further flare(s) occurred on the other side of the Sun.

 \subsection{Analysis of foot-point motions}

Let us now investigate the movement of opposite polarity sunspots before the solar eruptions in the selected clusters. In Figures \ref{12158map} and \ref{12192map}, we follow the area-weighted barycentres of the negative (square) and positive (triangle) polarities before the solar eruptions. We use the Carrington-coordinate system which is a coordinate system attached to the Sun. Richard C. Carrington determined the solar rotation rate by watching low-latitude sunspots in the 1850s. This Carrington-coordinate system rotates with the Sun in a sidereal frame exactly once every 25.38 days. Coordinates of the Carrington coordinate system are heliographic latitude ({\it B}) and heliographic longitude ({\it L}). The starting point of the Carrington frame of reference system is at noon (GMT) on 1st of January 1854. 

In Figures \ref{12158map}-\ref{12192map}, the {\it x}-axis is the heliographic longitude ({\it L}) and {\it y}-axis is the heliographic latitude ({\it B}). The colour bars demonstrate the evolution of the opposite polarity movements of sunspots. Black filled circles demonstrate the photospheric positions of the two area-weighted barycentres of polarities at the flare onset time.

Let us now investigate the motions of the area-weighted barycenters of flares with CME (AR 12158) and series of flares without CME (AR 12192). If we compare Figures \ref{12158map} and \ref{12192map} then we observe that the behaviour of the movements of opposite polarity sunspots of the  AR 12158  and the  AR 12192 are rather different before the eruptive events. In the case of a fast CME one barycenter has not moved around much (Fig.\ref{12158map}), while in the case of a series of flares with no CME both barycenters moved around, one even showing a remarkable and distinctive S-shape (see Fig. \ref{12192map}). Therefore, we will provide now further detailed analyses of the movements of opposite polarity of sunspots in the two AR cases. 

\begin{figure}[ht!]

\includegraphics{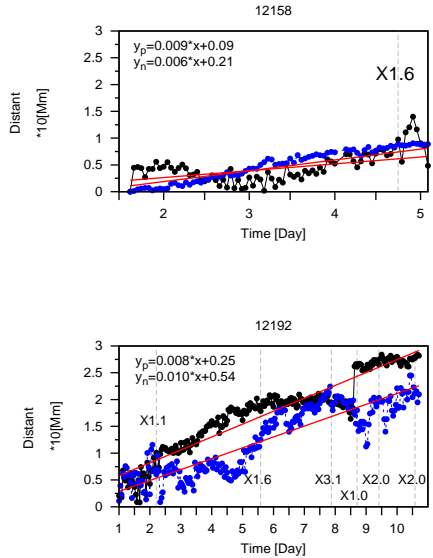}

\caption{The evolution of displacement from the reference position of positive (blue filled circles)/negative (black filled circles) polarity.  The red lines show the trend of  displacements in time. The upper figure is AR 12158 and lower panel is AR 12192. The {\it x}-axis is number of days measured from the reference time. The {\it y}-axis is displacement value. }
\label{distance}

\end{figure}

In the Figure \ref{distance}, we follow separately the evolution of negative (black filled circles) and positive polarity (blue filled circles) displacement as a function of time. The reference point is the first data point of the centres. We have calculated the distance between coordinates of the reference point and coordinates of each point as time progresses during the investigation. We have fitted a linear regressions to displacement values of the barycenters of the negative and positive polarities of the two ARs (denote by red lines in Figure \ref{distance}), respectively. 
The displacement of barycentre of the positive polarity of AR 12158 (see upper panel of Figure \ref{distance}) is $0.09^{\circ}$ per day but that of the negative polarity is  $0.2^{\circ}$.  These mean that the barycentre of the positive polarity is indeed staying very close to the reference point, but that of the negative polarity does not.
 The displacement of barycentre of the positive polarity of AR 12192 (see lower panel of Figure \ref{distance}) is $0.25^{\circ}$ and the negative polarity is  $0.55^{\circ}$ per day. Here, the two barycentres of polarity have moved around considerably, especially when compared to those of a flare eruption with an accompanying CME. We also notice the periodicity (both spatially, e.g. Fig. \ref{12192map}, and temporally, e.g. lower panel of Fig. \ref{distance}). Could the be distinct and deterministic cursors for fare eruption without CME? A preliminary insight (not shown here) indicates towards the positive answer. For a firm confirmation beyond a conjecture, another investigation is needed on a much larger database underpinned with a rigorous statistical analysis. 
We might conclude now that, for the selected clusters where we applied the $WG_{M}$ method, we found that: (i) when the AR (12158) produced flares with fast CME then the negative polarity has moved around before the fast CME occurred, and, the positive polarity sunspot ``stayed" at the same coordinates.Ê If, however, the AR (12192) produced flares without CME then the positive and the negative polarity sunspots {\it both} have moved around.Ê

\section{Results and Conclusion } \label{conclusion}

In this paper, we proposed the application of a relatively new measure of the pre-flare and pre-CME behaviour. We analysed the evolution of two typical active regions (AR 12158 and AR 12192) by using the SDO/HMI-Debrecen Data (HMIDD) sunspot catalogue. This two ARs individually represent situations where (i) the flare occurs with a fast CME (AR 12158) and (ii) there is (even more) high energetic flare eruptions without CME (AR 12192). The proxy measure of our approach is a so-called weighted horizontal gradient of magnetic field ($WG_{M}$) defined between spots of opposite polarities closer to the neutral line of an AR. The value and the temporal variation of $WG_{M}$ is found to possess novel, very interesting and potentially important diagnostic information about the intensity and onset time of expected flares, see \cite{Korsos2015}. 
In the  case of AR 12158  with a fast CME, the expected flare intensity by Eq. \ref{intensity} underestimates by almost an order of magnitude (the expected flare was M8.0, but the occurred was X1.6). If, however, we consider the standard error of the estimation, then, we will arrive at least at the expected flare intensity class rightly, i.e. there is X1.1-class which is closer to the actual occurred X1.6 flare intensity. On the other hand, in the case of AR 12192, we estimated the first expected flare intensity after the first maximum of $WG_{M}$ by Eq. \ref{intensity} as an X1.0-class flare and the occurred one was in fact X1.1-class. Next, we estimated the second expected flare intensity-class after the second maximum of $WG_{M}$ by Eq. \ref{intensity}  as an X1.4-class flare. Here, the flare intensity flare class is also well estimated because the AR 12192 produced a healthy series of X-class flares (X1.6, X3.1, X1.0, X2.0 and X2.0) after the second maximum of $WG_{M}$.

We have also predicted flare onset times by means of Equation \ref{time}. We have used the measured duration of the converging phase ($T_{C}$) of motion of opposite polarities for the prediction of flare onset time. We have recognised two converging-diverging motion before the X1.6 flare that occurred with CME from AR 12158. There was only 5 hrs between the predicted onset time and the real occurrence time of the X1.6-class flare. Here, we note that after the first converging-diverging motion a flare may failed to occur, and only after the second converging-diverging motion there was a flare. Two consecutive converging-diverging motions may be a pre-CME diagnostic tool, but we need more flares with fast CME occurrence examples for a further investigation and definite conclusion.

Next, we investigated the predicted flare(s) onset time of AR 12192. We followed the evolution of the distance parameter in time. First, we have identified the duration of converging and diverging motion of barycenters of opposite polarities  before the first X1.1-class eruption. Second, we also can see a further converging and diverging motion subsequently followed by X1.6, X3.1, X1.0, X2.0 and X2.0 class flares. The first predicted onset time of the X1.1 flare is 10 hrs earlier than the actual observed flare onset time. In the next step, we estimated the next expected flare onset time from the second approaching and receding phases of the distance before the series of X1.6, X3.1, X1.0, X2.0 and X2.0-class energetic flares. The second predicted onset time of the next flare was at 02:00 on 21 September 2014, obtained by Equation \ref{time}, and the first flare (X1.6) occurred 12 hrs later after the second converging-diverging motion. Again, there seems to be an underestimate of onset time. Note that we cannot predict onset time of the subsequent flares, but we used the percentage value of the maximum of $WG_{M}$ to the value of $WG_{M}$ at the flare onset for the prediction of intensity of the next X3.1, X1.0, X2.0 and X2.0-class energetic flares.

In  \cite{Korsos2015}, we proposed an empirical percentage that of $WG_{M}$ that is a proxy of the available non-potential energy to be released in a spot group. We determine this percentage value from the relationship of the maximum of $WG_{M}$ and the value of $WG_{M}$ at the flare onset. In the case of AR 12192, the percentage of decrease is 28\% at the X1.1 flare onset, so more flare is expected. Next, after the second maximum value of the $WG_{M}$, in the case of the X1.6-class flare, the percentage is only 20\% and indeed more flares (X3.1, X1.6, X1.0, X2.0 and X2.0) occurred during the decreasing phase of the $WG_{M}$. The percentage of the maximum of $WG_{M}$ and the value of $WG_{M}$ at the last X2.0-class energetic flare is 38\% but no more flares seemed to occur during the decreasing phase of the $WG_{M}$. Here, we note that the AR 12192 turned out off the west limb  becoming invisible.

Perhaps the most interesting results of the present case studies are the movements (i.e. loci) of the area-weighted barycenters of the opposite polarity sunspots before flare and CME in the selected clusters. We have used the Carrington-coordinate system which is a fixed solar coordinate system to clearly follow the evolution of the movements of barycenters of the opposite polarity sunspots. In particular, Figures \ref{12158map} and \ref{12192map} reveal a very interesting feature: they capture some distinct patterns of the behavior of the negative and the positive polarity sunspots before flares. There are two types of figures: one for the case with fast CME (see Figure \ref{12158map}), another flare(s) without CME (see Figure \ref{12192map}). Even just a simple visual inspection of the trajectory of sunspot barycenter motion will unveil that one may find {\it distinct signatures} of flaring without fast CMEs, respectively, in the trajectories of sunspots. Furthermore, we have determined separately the evolution of negative and positive polarity displacement, i.e. as a function of time of the two ARs (AR 12158 and AR 12192). We have found, where there is a fast CME the negative polarity area-weighted barycenter has moved before the fast CME and the positive polarity ``stayed" at the same coordinates. Otherwise, when the AR produced flares but no CME, then the barycenters of {\it both} the negative and the positive polarity sunspots have moved around. This latter empirical relation may mean that the highly stressed region would relax itself via these unwinding motions, not leaving enough free energy there for a major mass uplift. The results are encouraging but we need to confirm statistically this latter statement by carrying out the study on much larger samples.

    \section{ Acknowledgements} 
        MBK is grateful to the University of Sheffield for the support received while carrying out research for some time there.  RE and MSR are grateful to Science and Technology Facilities Council (STFC) UK and the Royal Society (UK). The authors also acknowledge the support received from the CAS Key Laboratory of Solar Activity, National Astronomical Observatories  Commission for Collaborating Research Program. RE acknowledges the support received from the CAS Presidents International Fellowship Initiative, Grant No. 2016VMA045.

\end{document}